\documentclass[conference,a4paper]{IEEEtran}

\usepackage{amssymb, amsmath, color}
\usepackage{amsthm,graphicx,enumerate,verbatim,xcolor}
\usepackage[small]{caption}
\usepackage{subfigure}
\usepackage{tikz}
\usetikzlibrary{arrows,positioning,fit,backgrounds}
\usetikzlibrary{calc}
\usepackage{bbm}
\usepackage{mathtools}
\usepackage{epstopdf}
\usepackage{float}

\newtheorem{thm}{Theorem}
\newtheorem{cor}{Corollary}
\newtheorem{lem}{Lemma}
\newtheorem{defn}{Definition}

\theoremstyle{remark}

\def\squarebox#1{\hbox to #1{\hfill\vbox to #1{\vfill}}}
\newcommand{\inout}{-}
\newcommand{\Pbar}{\overline{P}}

\newcommand{\Ebb}{\mathbb{E}}

\newcommand{\Pbb}{\mathbb{P}}

\newcommand{\Ccal}{\mathcal{C}}

\newcommand{\Mcal}{\mathcal{M}}

\newcommand{\Scal}{\mathcal{S}}

\newcommand{\Ucal}{\mathcal{U}}

\newcommand{\Xcal}{\mathcal{X}}
\newcommand{\Ycal}{\mathcal{Y}}

\newcommand{\Lcal}{\mathcal{L}}
\newcommand{\cn}{\mathcal{C}^n}
\newcommand{\Pbf}{\mathbf{P}}
\newcommand{\Qbf}{\mathbf{Q}}
\newcommand{\Scheck}{\check{S}}

\newcommand{\Shat}{\hat{S}}

\begin{document}

\sloppy

\title{Joint Source-Channel Secrecy Using Hybrid Coding}


\author{Eva C. Song \qquad Paul Cuff \qquad H. Vincent  Poor\\ Dept. of Electrical Eng., Princeton University,  NJ 08544\\ \{csong, cuff, poor\}@princeton.edu}


\maketitle
\begin{abstract}
The secrecy performance of a source-channel model is studied in the context of lossy source compression over a noisy broadcast channel. The source is causally revealed to the eavesdropper during decoding. The fidelity of the transmission to the legitimate receiver and the secrecy performance at the eavesdropper are both measured by a distortion metric. Two achievability schemes using the technique of hybrid coding are analyzed and compared with an operationally separate source-channel coding scheme. A numerical example is provided and the comparison results show that the hybrid coding schemes outperform the operationally separate scheme.
\end{abstract}
\section{Introduction} \label{intro}
The secrecy properties of the wiretap channel have been studied under a variety of formulations. Shannon \cite{shannon} considered the model of a noiseless wiretap channel where the transmitter and the legitimate receiver share a secret key. Other works \cite{wyner} \cite{ck} consider the case of a noisy wiretap channel where the physical structure of the channel is exploited instead of using a secret key. 

The most frequently adopted measure for information theoretic secrecy in the literature by far is equivocation, or normalized equivocation to be more precise. Wyner \cite{wyner} introduced the notion of (normalized) equivocation for the study of secrecy capacity of a wiretap channel. This secrecy metric uses the conditional entropy of the source given what the eavesdropper observes $H(S|E)$, where $E$ here can be a noisy channel output or an ciphered text protected by a secret key depending on the setup. When the source is a sequence, this quantity is typically normalized over the blocklength, $\frac1nH(S^n|E)$. Such a metric can be intuitively interpreted as the average statistical independence between the source and what the eavesdropper observes.

Inspired by \cite{yamamoto}, other works \cite{cuff-allerton} \cite{cuff-globecom} \cite{schieler-isit12} \cite{song-tcom} \cite{song-allerton14} have taken a rate-distortion approach to secrecy in communication systems. Instead of using equivocation, the secrecy is measured by the average distortion between the source and the eavesdropper's reconstruction of the source by allowing the eavesdropper to optimize its estimation. There the goal is to design an encoding and decoding scheme such that the source can be delivered reliably (lossless or lossy) to the legitimate receiver while a high distortion can be forced on the eavesdropper.

It may not have been straightforward to draw any connection between these two secrecy metrics until recent work \cite{schieler-rd} which has shown that equivocation is a special case of the distortion secrecy metric if the source sequence realization is causally disclosed to the eavesdropper during decoding. Specifically, the distortion secrecy formulation with causal source disclosure fully recovers the equivocation secrecy formulation by choosing the distortion function to be a log-loss function. 

In this work, we investigate the secrecy performance of a source-channel communication system composed of an independent and identically distributed (i.i.d.) source sequence and a noisy memoryless wiretap channel. By causally disclosing the source to the eavesdropper and using the distortion secrecy metric, it grants us the freedom of considering the general formulation of a secrecy problem, which can be particularized to the equivocation formulation if needed. A variation of this source-channel secrecy model was considered in \cite{song-tcom} without causal source disclosure. Despite an important game-theoretic setting, such formulation does not render a strong enough secrecy criterion.

Previous work \cite{separate-schieler} considers the same source-channel model with causal source disclosure. However, only an operationally separate source-channel coding scheme was considered. Recent work on hybrid coding \cite{hybrid} and the likelihood encoder \cite{cuff-itw2013} \cite{song-isit2014} suggests a new way of approaching this problem.
\vspace{-0.2cm}
\section{Preliminaries} \label{prelim}
\vspace{-0.2cm}
\subsection{Notation} \label{notation}
A sequence $X_1,..., X_n$ is denoted by $X^n$. Limits taken with respect to ``$n\rightarrow \infty$" are abbreviated as ``$\rightarrow_n$". Inequalities of the forms $\limsup_{n\rightarrow \infty}h_n\leq h$ and $\liminf_{n\rightarrow \infty}h_n\geq h$ are abbreviated as $h_n\leq_n h$ and $h_n\geq_n h$, respectively. 
When $X$ denotes a random variable, $x$ is used to denote a realization, and $\mathcal{X}$ is used to denote the support of that random variable.  A Markov relation is denoted by the symbol $\inout$. 
We use $\Ebb_P$, $\Pbb_P$, and $I_{P}(X;Y)$ to indicate expectation, probability, and mutual information taken with respect to a distribution $P$; however, when the distribution is clear from the context, the subscript will be omitted. We use a bold capital letter $\mathbf{P}$ to denote that a distribution $P$ is random. 

\enlargethispage*{\baselineskip}
For a distortion measure $d: \mathcal{X} \times \mathcal{Y}\mapsto \mathbb{R}^+$, we use $\Ebb \left[d(X,Y)\right]$ to measure the distortion of $X$ incurred by reconstructing it as $Y$. The maximum distortion is defined as
$$d_{max}=\max_{(x,y) \in \Xcal\times\Ycal} d(x,y).$$
The distortion between two sequences is defined to be the per-letter average distortion 
$$d(x^n,y^n)=\frac1n\sum_{t=1}^n d(x_t,y_t).$$

\subsection{Total Variation Distance}
The total variation distance between two probability measures $P$ and $Q$ on the same $\sigma$-algebra $\mathcal{F}$ of subsets of the sample space $\Xcal$ is defined as
$$\lVert P-Q\rVert_{TV}\triangleq \sup_{\mathcal{A}\in \mathcal{F}}|P(\mathcal{A})-Q(\mathcal{A})|.$$

Some basic properties of total variation distance are given as Property 2 in \cite{schieler}. 
\vspace{-0.1in}
\subsection{Soft-covering Lemma}
We now introduce the soft-covering lemma, which will be used in the achievability proof of the joint source-channel coding scheme. 
\begin{lem}(\textbf{Soft-covering}, \cite{cuff2012distributed}) \label{bsc}
Given a joint distribution $\Pbar_{UXZ}$, let $\Ccal^{(n)}$ be a random codebook of sequences $U^n(m)$, with $m=1,...,2^{nR}$, each drawn independently and i.i.d. according to $\Pbar_U$. Let
\vspace{-0.1in}
\begin{eqnarray}
&&\Pbf_{MX^nZ^k}(m,x^n,z^k)\nonumber\\
&\triangleq& \frac{1}{2^{nR}}\prod_{t=1}^n \Pbar_{X|U}(x_t|U_t(m))\prod_{t=1}^k\Pbar_{Z|XU}(z_t|x_t,U_t(m))
\end{eqnarray}
\vspace{-0.1in}
and 
\vspace{-0.1in}
\begin{eqnarray}
\Pbar_{X^nZ^k}\triangleq\prod_{t=1}^n\Pbar_X(x_t)\prod_{t=1}^k\Pbar_{Z|X}(z_t|x_t).
\end{eqnarray}
If $R>I(X;U)$, then
$$\Ebb_{\cn}\left[\left\Vert \Pbf_{X^nZ^k}-\Pbar_{X^nZ^k}\right\Vert_{TV}\right]\leq \exp(-\gamma n)\rightarrow_n 0,$$
for any $\beta< \frac{R-I(X;U)}{I(Z;U|X)}$, $k\leq \beta n$, where $\gamma>0$ depends on the gap $\frac{R-I(X;U)}{I(Z;U|X)}-\beta$.
\end{lem}
\section{Problem Setup and Previous Work}
\subsection{Problem Setup} \label{setup}
Given a memoryless source and broadcast channel, we want to maximize the distortion forced on the eavesdropper (for estimating the source) while communicating the source reliably within a distortion constraint to the legitimate receiver.
The input of the system is an i.i.d. source sequence $S^n$ distributed according to $\prod_{t=1}^n\Pbar_S(s_t)$ and the channel is a memoryless broadcast channel $\prod_{t=1}^n \Pbar_{YZ|X}(y_t,z_t|x_t)$. The source realization is causally disclosed to the eavesdropper during decoding. The source-channel coding model satisfies the following constraints:
\begin{itemize}
\item Encoder $f_n: \Scal^n \mapsto \Xcal^n$ (possibly stochastic);
\item Legitimate receiver decoder $g_n: \Ycal^n \mapsto \hat{\Scal}^n$ (possibly stochastic);
\item Eavesdropper decoders $\{P_{\Scheck_t|Z^nS^{t-1}}\}_{t=1}^n$.
\end{itemize}
The system performance is measured by a distortion metric $d(\cdot,\cdot)$ as follows:
\begin{itemize}
\item Average distortion for the legitimate receiver: 
$$\Ebb\left[d(S^n,\Shat^n)\right]\leq_n D_b$$
\item Minimum average distortion for the eavesdropper:
$$\min_{\{P_{\Scheck_t|Z^nS^{t-1}}\}_{t=1}^n}\Ebb[d(S^n,\Scheck^n)]\geq_n D_e$$
\end{itemize}

\begin{defn}
A distortion pair $(D_b,D_e)$ is achievable if there exists a sequence of source-channel encoders and decoders $(f_n,g_n) $ such that 
$$\Ebb[d(S^n,\Shat^n)]\leq_n D_b$$
and 
$$\min_{\{P_{\Scheck_t|Z^nS^{t-1}}\}_{t=1}^n}\Ebb[d(S^n,\Scheck^n)]\geq_n D_e.$$
\end{defn}

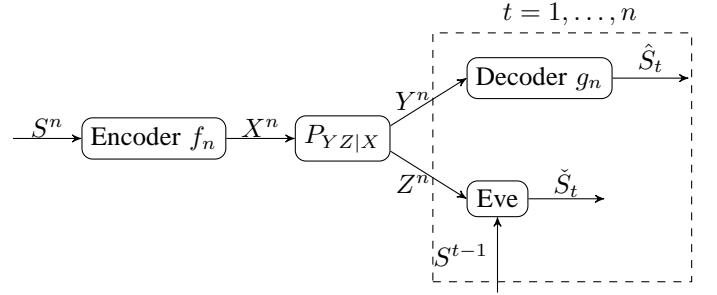
\begin{figure}[H]
\begin{tikzpicture}
[node distance=1cm,minimum height=4mm,minimum width=8mm,arw/.style={->,>=stealth'}]
  \node[coordinate] (source) {};
  \node[rectangle,draw,rounded corners] (alice) [right =9mm of source] {Encoder $f_n$};
  \node[rectangle,draw,rounded corners] (ch) [right =9mm of alice] {$P_{YZ|X}$};
  \node[rectangle,draw,rounded corners] (bob) [right =of ch,yshift=8mm] {Decoder $g_n$};
  \node[rectangle,draw,rounded corners] (eve) [right =of ch,yshift=-8mm] {Eve};
  \node[coordinate] (shat) [right =of bob] {};
  \node[coordinate] (t) [right =of eve] {};
  \node[rectangle] at ([xshift=4mm,yshift=8.5mm] bob.center) {$t=1,\ldots,n$};
  \node[coordinate] (past) [below =of eve] {};

  \draw [arw] (source) to node[midway,above,yshift=-1mm]{$S^n$} (alice);
  \draw [arw] (alice) to node[midway,above,yshift=-1mm]{$X^n$} (ch);
  \draw [arw] (ch.15) to node[pos=0.3,above,yshift=-1mm]{$Y^n$} (bob.west);
  \draw [arw] (ch.345) to node[pos=0.3,below]{$Z^n$} (eve.west);
  \draw [arw] (bob) to node[midway,above,yshift=-.5mm]{$\hat{S}_t$} (shat);
  \draw [arw] (eve) to node[midway,above,yshift=-1mm]{$\Scheck_t$} (t);
  \draw [arw] (past) to node [midway,left] {${S}^{t-1}$} (eve); 
  \draw [dashed] ([xshift=-0.85cm,yshift=-11mm] eve.center) rectangle ([xshift=2.0cm,yshift=6mm] bob.center);
\end{tikzpicture}
\caption{Joint source-channel secrecy system setup with causal source disclosure at the eavesdropper}
\label{source_channel_fig}
\end{figure}

The above mathematical formulation is illustrated in Fig. \ref{source_channel_fig}.
\vspace{-0.5cm}
\subsection{Previous Result}
Before introducing the new joint source-channel coding schemes, we first review the achievability result from previous work \cite{separate-schieler} of the same problem formulation with an operationally separate source-channel coding scheme. Although the problem was only studied for the case of lossless reconstruction at the legitimate receiver in \cite{separate-schieler}, the result can be readily generalized to the case of lossy compression as was formulated in Section \ref{setup}. 

\begin{thm}(Generalized Theorem 2 of \cite{separate-schieler}) \label{thm-separate}
A distortion pair $(D_b, D_e)$ is achievable if 
\begin{eqnarray}
I(S;U_1)<I(U_2;Y)\ \ \ \ \ \ \ \ \ \ \ \ \ \ \ \ \ \ \ \ \ \ \ \ \ \ \ \ \ \label{thm1-1}\\
I(S;\Shat|U_1)<I(V_2;Y|U_2)-I(V_2;Z|U_2)\ \ \ \ \ \ \ \ \label{thm1-2}\\
D_b\geq \Ebb\left[d(S,\Shat)\right]\ \ \ \ \ \ \ \ \ \ \ \ \ \ \ \ \ \ \ \ \ \ \ \ \  \label{thm1-3}\\
D_e\leq \eta \min_{a\in \hat{\Scal}}\Ebb[d(S, a)]+(1-\eta)\min_{t(u_1)}\Ebb[d(S,t(U_1))]\label{thm1-4}
\end{eqnarray}
for some distribution $\Pbar_S\Pbar_{\Shat|S}\Pbar_{U_1|\Shat}\Pbar_{U_2}\Pbar_{V_2|U_2}\Pbar_{X|V_2}\Pbar_{YZ|X}$, where 
\begin{eqnarray}
\eta=\frac{[I(U_2;Y)-I(U_2;Z)]^+}{I(S;U_1)}.
\end{eqnarray}
\end{thm}

Since the source coding and channel coding parts of the above scheme are almost independent (with some technical details), we refer to it as the operationally separate source-channel coding scheme -- Scheme O. 

\section{Main Results}
This section is organized as follows. We first introduce the idea of secure hybrid coding. We then state the result of using basic hybrid coding (Scheme I) followed by its proof. We next state and briefly discuss the result using superposition hybrid coding (Scheme II). Finally, we analytically compare Scheme O, I and II, and give a trivial outer bound for completeness.
\vspace{-0.2cm}
\subsection{Secure Hybrid Coding}
Hybrid coding is a joint source-channel coding technique \cite{hybrid} where 
1) the encoder generates a digital codeword from the analog source and selects the channel input as a symbol-by-symbol function of the codeword and the source; and 2) the decoder recovers the digital codeword from the analog channel output and selects the source estimate as a symbol-by-symbol function of the codeword and the channel output. It has been shown that this joint source-channel code is at least optimal for point-to-point communication. For the purpose of achieving secrecy, the symbol-by-symbol mapping (deterministic) to the channel input in the encoding stage is modified to be stochastic. 
\vspace{-0.2cm}
\subsection{Scheme I -- Basic Hybrid Coding}
An achievability region using basic secure hybrid coding is given in the following theorem.
\begin{thm} \label{achievability-single}
A distortion pair $(D_b, D_e)$ is achievable if 
\begin{eqnarray}
I(U;S)&<&I(U;Y)\\
D_b&\geq&\Ebb[d(S,\phi(U,Y))]\\
D_e&\leq&\beta \min_{\psi_0(z)}\Ebb[d(S,\psi_0(Z))]\nonumber\\
&&+(1-\beta)\min_{\psi_1(u,z)}\Ebb[d(S,\psi_1(U,Z))]
\end{eqnarray}
where
\begin{eqnarray}
\beta=\min\left\{\frac{[I(U;Y)-I(U;Z)]^+}{I(S;U|Z)},1\right\}
\end{eqnarray}
for some distribution $\Pbar_S\Pbar_{U|S}\Pbar_{X|SU}\Pbar_{YZ|X}$ and function $\phi(\cdot,\cdot)$.
\end{thm}

The proof of Theorem \ref{achievability-single} to be presented next uses hybrid coding combined with the likelihood encoder. The general idea is that under our choice of the encoder and decoder, the system induced distribution $\Pbf$ is close in total variation distance to an idealized distribution $\Qbf$ by our construction. Therefore, by properties of total variation, we can approximate the performance of the system under $\Pbf$ by that under $\Qbf$.
\enlargethispage*{\baselineskip}
\subsection{Proof Outline of Scheme I}
The source and channel distributions $\Pbar_S$ and $\Pbar_{YZ|X}$ are given by the problem statement. Fix a joint distribution $\Pbar_S\Pbar_{U|S}\Pbar_{X|SU}\Pbar_{YZ|X}$. We will use $\Pbar_{S^n}$ to denote $\prod_{t=1}^n\Pbar_S$.

\textbf{Codebook generation:} We independently generate $2^{nR}$ sequences in $\Ucal^n$ according to $\prod_{t=1}^n\Pbar_U(u_t)$ and index by $m\in[1:2^{nR}]$. We use $\Ccal^{(n)}$ to denote this random codebook. 

\textbf{Encoder:} Encoding has two steps. 
In the first step, a likelihood encoder $\Pbf_{LE}(m|s^n)$ is used. It chooses $M$ stochastically according to the following probability:
\begin{eqnarray}
\Pbf_{LE}(m|s^n)=\frac{\Lcal(m|s^n)}{\sum_{\bar{m}\in\Mcal}\Lcal(\bar{m}|s^n)}
\end{eqnarray}
where $\Mcal=[1:2^{nR}]$, and 
\begin{eqnarray}
\Lcal(m|s^n)=\Pbar_{S^n|U^n}(s^n|u^n(m)).
\end{eqnarray}
In the second step, the encoder produces the channel input through a random transformation given by 
$\prod_{t=1}^n\Pbar_{X|SU}(x_t|s_t,U_t(m)).$
\newpage
\textbf{Decoder:} Decoding also has two steps.
In the first step, let $\Pbf_{D1}(\hat{m}|y^n)$ be a good channel decoder with respect to the  codebook $\{u^n(a)\}_{a}$ and memoryless channel $\Pbar_{Y|U}$. In the second step, fix a function $\phi(\cdot,\cdot)$. Define $\phi^n(u^n,y^n)$ as the concatenation $\{\phi(u_t,y_t)\}_{t=1}^n$ and set the decoder $\Pbf_{D2}$ to be the deterministic function
\begin{eqnarray}
\Pbf_{D2}(\hat{s}^n|\hat{m},y^n)\triangleq\mathbbm{1}\{\hat{s}^n=\phi^n(u^n(\hat{m}),y^n)\}.
\end{eqnarray}

\textbf{Analysis:}
We can write the system induced distribution in the following form:
\begin{eqnarray}
&&\Pbf_{MU^nS^nX^nY^nZ^n\hat{M}\hat{S}^n}(m,u^n,s^n,x^n,y^n,z^n,\hat{m},\hat{s}^n)\nonumber\\
&\triangleq&\Pbar_{S^n}(s^n){\Pbf_{LE}(m|s^n)\mathbbm{1}\{u^n=U^n(m)\}}\nonumber\\
&&{\prod_{t=1}^n \Pbar_{X|SU}(x_t|s_t,u_t)}\prod_{t=1}^n \Pbar_{YZ|X}(y_t,z_t|x_t)\nonumber\\
&&{\Pbf_{D1}(\hat{m}|y^n)\Pbf_{D2}(\hat{s}^n|\hat{m},y^n)}.
\end{eqnarray}

An idealized distribution $\Qbf$ is defined as follows to help with the analysis:
\begin{eqnarray}
&&\Qbf_{MU^nS^nX^nY^nZ^n}(m,u^n,s^n,x^n,y^n,z^n)\nonumber\\
&\triangleq&\frac{1}{2^{nR}}\mathbbm{1}\{u^n=U^n(m)\}\prod_{t=1}^n\Pbar_{S|U}(s_t|u_t)\nonumber\\
&&\prod_{t=1}^n\Pbar_{X|SU}(x_t|s_t,u_t)\prod_{t=1}^n\Pbar_{YZ|X}(y_t,z_t|x_t).
\end{eqnarray}

\subsubsection{Distortion analysis at the legitimate receiver}
Applying Lemma \ref{bsc} and properties of total variation distance, if 
\begin{eqnarray}
{R>I(U;S)}, \label{sr}
\end{eqnarray} 
then
\begin{eqnarray}
\Ebb_{\Ccal^{(n)}}\left[\left\Vert\Pbf-\Qbf\right\Vert_{TV}\right]\leq \exp(-\gamma_1 n)\triangleq {\epsilon_1}_n \rightarrow_n 0, \label{P2Q}
\end{eqnarray}
where the distributions are over the random variables $MU^nS^nX^nY^nZ^n$,

Using the same steps as was given in \cite{song-isit2014} for the analysis of the Wyner-Ziv setting, it can be verified that the following holds:
\begin{eqnarray}
&&\Ebb_{\Ccal^{(n)}}\left[\Ebb_{\Pbf}\left[d(S^n,\hat{S}^n)\right]\right]\nonumber\\
&\leq&\Ebb_{\Pbar}[d(S,\phi(U,Y))]+d_{max}({\epsilon_1}_n+\delta_n), \label{aveD}
\end{eqnarray}
if
\begin{eqnarray}
{R\leq I(U;Y)}, \label{cr}
\end{eqnarray}
where $\delta_n\rightarrow_n 0.$

\subsubsection{Distortion analysis at the eavesdropper}
On the eavesdropper side, we make the following observation. Define an auxiliary distribution 
\begin{eqnarray}
\check{\Qbf}^{(i)}_{S^i Z^n}(s^i,z^n)
\triangleq\prod_{t=1}^n\Pbar_Z(z_t)\prod_{j=1}^i\Pbar_{S|Z}(s_j|z_j). \label{Qc}
\end{eqnarray}
Under $\check{\Qbf}^{(i)}$,
\begin{eqnarray}
S_i\inout Z_i\inout Z^nS^{i-1}. \label{markov1}
\end{eqnarray}
Recall that 
\begin{eqnarray}
\Qbf_{MZ^nS^i}(m,z^n,s^i)\ \ \ \ \ \ \ \ \ \ \ \ \ \ \ \ \ \ \ \ \ \ \ \ \ \ \ \ \ \ \ \ \ \ \ \ \ \nonumber\\
=\frac{1}{2^{nR}}\prod_{t=1}^n\Pbar_{Z|U}(z_t|U_t(m))\prod_{j=1}^i\Pbar_{S|ZU}(s_j|z_j,U_j(m))
\end{eqnarray}
\normalsize
and under ${\Qbf}$, the following Markov relation holds:
\begin{eqnarray}
S_i\inout Z_iU_i(M)\inout Z^nS^{i-1}M. \label{markov2}
\end{eqnarray}

Applying Lemma \ref{bsc}, we have
\begin{eqnarray}
\Ebb_{\Ccal^{(n)}}\left[\left\Vert\check{\Qbf}^{(i)}_{Z^nS^i}-{\Qbf}_{Z^nS^i}\right\Vert_{TV}\right]\leq \exp(-\gamma_2 n)
\end{eqnarray}
if 
\begin{eqnarray}
R>I(Z;U) \label{pieq6}
\end{eqnarray}
where $i$ here can go up to $\beta n$, for any $\beta<\frac{R-I(U;Z)}{I(S;U|Z)}.$ Consequently,
\small
\begin{eqnarray}
\Ebb_{\Ccal^{(n)}}\left[\left\Vert\check{\Qbf}^{(i)}_{Z^nS^i}-\Pbf_{Z^nS^i}\right\Vert_{TV}\right]
\leq \exp(-\gamma_1 n)+\exp(-\gamma_2 n). \label{P2Qc}
\end{eqnarray}
\normalsize
Note that $(\ref{pieq6})$ is a degenerate statement if $R>I(Z;U)$. 

Also note that since $R>0$, we have
\begin{eqnarray}
\Ebb_{\Ccal^{(n)}}\left[\left\Vert\Qbf_{u_i(M)}-\Pbar_U\right\Vert_{TV}\right]\leq \exp(-\gamma_3 n). \label{Q2Pb}
\end{eqnarray}

Therefore, combining $(\ref{P2Q})$, $(\ref{P2Qc})$, $(\ref{Q2Pb})$, and $(\ref{aveD})$, there exists a codebook $\Ccal^{(n)}$ such that 
\begin{eqnarray}
\sum_{i=1}^n \left\Vert P_{MZ^nS^i}-{Q}_{MZ^nS^i}\right\Vert_{TV}\leq \epsilon_n \label{pP2Qt}\\
\sum_{i=1}^n \left\Vert P_{Z^nS^i}-\check{Q}^{(i)}_{Z^nS^i}\right\Vert_{TV}\leq \epsilon_n \label{pP2Qc}\\
\sum_{i=1}^n \left\Vert{Q}_{u_i(M)}-\Pbar_{U}\right\Vert_{TV}\leq \epsilon_n \label{pQt2Pb}\\
\Ebb_{P} \left[d(S^n,\hat{S}^n)\right]\leq \Ebb_{\Pbar}\left[d(S^n,\hat{S}^n)\right]+\epsilon_n \label{pdP}
\end{eqnarray}
where 
$\epsilon_n=n\left(2\exp(-n\gamma_1)+\exp(-n\gamma_2)+\exp(-n\gamma_3)\right)
+d_{max}({\epsilon_1}_n+\delta_n)\rightarrow_n 0.$

Now we can bound the distortion at the eavesdropper by breaking it down into two sections. The distortion after the time transition $\beta n$ can be lower bounded by the following: 
%
%
\small
\begin{eqnarray}
&&\min_{\{{\psi_1}_i(s^{i-1},z^n)\}}\Ebb_{P}\left[\frac1k\sum_{i=j}^n d(S_i, {\psi_1}_i(S^{i-1},Z^n))\right]\nonumber\\
&=&\frac1k \sum_{i=j}^n\min_{{\psi_1}_i(s^{i-1},z^n)}\Ebb_{P}\left[d(S_i,{\psi_1}_i(S^{i-1},Z^n))\right]\\
&\geq&\frac1k \sum_{i=j}^n \min_{{\psi_1}_i(s^{i-1},z^n,m)}\Ebb_{P}\left[d(S_i, {\psi_1}_i(S^{i-1},Z^n,M))\right]\\
&\geq&\frac1k \sum_{i=j}^n \min_{{\psi_1}_i(s^{i-1},z^n,m)}\Ebb_{{Q}}\left[d(S_i, {\psi_1}_i(S^{i-1},Z^n,M))\right]\nonumber\\
&&-\epsilon_n d_{max}\label{pdQt}\\
&=& \frac1k \sum_{i=j}^n \min_{\psi_1(u,z)}\Ebb_{{Q}} \left[d(S_i,\psi_1(u_i(M),Z_i))\right]-\epsilon_n d_{max}\label{pdQtu}\\
&\geq&\frac1k \sum_{i=j}^n \min_{\psi_1(u,z)} \Ebb_{\Pbar}\left[d(S,\psi_1(U,Z))\right]-2\epsilon_n d_{max} \label{pdQt2Pb}
\end{eqnarray}
\normalsize
where $k=(1-\beta) n$, $j=\beta n+1$, $(\ref{pdQt})$ is from $(\ref{pP2Qt})$, $(\ref{pdQtu})$ uses the Markov relation given in $(\ref{markov2})$, and $(\ref{pdQt2Pb})$ uses $(\ref{pQt2Pb})$ and the fact that 
$${Q}_{Z_iS_i|U_i}(z_i,s_i|u_i)=\Pbar_{Z|U}(z_i|u_i)\Pbar_{S|ZU}(s_i|z_i,u_i).$$
Similarly, by repeating the above process by replacing $Q$ with $\check{Q}$ using $(\ref{pP2Qc})$, the Markov relation given in $(\ref{markov1})$, and the definition of $\check{\Qbf}$ given in $(\ref{Qc})$, we can lower bound the distortion before time $\beta n$ as 
\begin{eqnarray}
&&\min_{\{{\psi_0}_i(s^{i-1},z^n)\}_i} \Ebb_P\left[\frac1k \sum_{i=1}^k d(S_i,{\psi_0}_i(S^{i-1},Z^n))\right]\nonumber\\
&\geq& \frac1k \sum_{i=1}^k\min_{\psi_0(z)}\Ebb_{\Pbar}\left[d(S,\psi_0(Z))\right]-\epsilon_n d_{max},\label{m3}
\end{eqnarray}
where $k=\beta n$.
Collecting $(\ref{sr})$, $(\ref{cr})$, and $(\ref{pdP})$ and taking the average of the distortion at the eavesdropper over the entire blocklength $n$ from $(\ref{pdQt2Pb})$ and $(\ref{m3})$ finishes the proof. \qed

\subsection{Scheme II -- Superposition Hybrid Coding}
An achievability region using superposition secure hybrid coding is given in the following theorem.
\begin{thm} \label{achievability}
A distortion pair $(D_b, D_e)$ is achievable if 
\begin{eqnarray}
I(V;S)&<&I(UV;Y)\label{r1}\\
D_b&\geq&\Ebb\left[d(S,\phi(V,Y))\right]\label{db}\\
D_e&\leq&\min\{\beta,\alpha\} \min_{\psi_0(z)}\Ebb\left[d(S,\psi_0(Z))\right]\nonumber\\
&&+\left(\alpha-\min\{\beta,\alpha\}\right)\min_{\psi_1(u,z)}\Ebb\left[d(S,\psi_1(U,Z))\right]\nonumber\\
&&+(1-\alpha)\min_{\psi_2(v,z)}\Ebb\left[d(S,\psi_2(V,Z))\right] \label{de} 
\end{eqnarray}
where
\begin{eqnarray}
\beta&=&\min\left\{\frac{[I(U;Y)-I(U;Z)]^+}{I(S;U|Z)},1\right\}\label{al}\\
\alpha&=&\min\left\{\frac{[r_s-I(Z;V|U)]^+}{I(S;V|ZU)},1\right\}\label{be}\\
r_s&=&\min\{I(V;Y|U),I(UV;Y)-I(S;U)\}\label{rs}
\end{eqnarray}
for some distribution $\Pbar_S\Pbar_{V|S}\Pbar_{U|V}\Pbar_{X|SUV}\Pbar_{YZ|X}$ and function $\phi(\cdot,\cdot)$.
\end{thm}

The proof of Theorem \ref{achievability} follows the same line as the proof of Theorem \ref{achievability-single} with the modification of using a superposition codebook and the superposition version of the soft-covering lemma which was discussed in Corollary VII.8 of \cite{cuff2012distributed}.

Under Scheme II, the distortion at the eavesdropper can potentially experience two transitions at $\beta n$ and $\alpha n$ due to the superposition structure of the code. 
\vspace{-0.08in}
\subsection{Scheme Comparison}
The relationships among Scheme O, I and II can be summarized in the following corollaries.
\begin{cor}\label{cor1}
Scheme II generalizes Scheme I.
\end{cor}
To see this, notice that we can let $U=\varnothing$ in Theorem \ref{achievability}. In fact, Scheme II simplifies to Scheme I if $\beta\geq\alpha$.
\begin{cor}\label{cor2}
Scheme O is a special case of Scheme II.
\end{cor}
This can be verified by using the following assignment of random variables from Theorem \ref{thm-separate} to \ref{achievability}:
$$U\leftarrow U_1U_2 \text{ and }V\leftarrow \Shat V_2$$
to show that the inequalities $(\ref{thm1-1})$ and $(\ref{thm1-2})$ satisfy the inequality $(\ref{r1})$, $\beta=\eta$, and $\alpha=1$. The equivalence of $(\ref{thm1-3})$ and $(\ref{thm1-4})$ to $(\ref{db})$ and $(\ref{de})$ can be obtained by using the statistical independence of $S\Shat U_1$ and $U_2V_2YZ$.
%
%
%

\subsection{The Perfect Secrecy Outer Bound}
\begin{thm}
If $(D_b, D_e)$ is achievable, then 
\begin{eqnarray}
I(S;U)&\leq&I(U;Y)\label{scp2pr}\\
D_b&\geq&\Ebb[d(S,\phi(U,Y))]\label{scp2pd}\\
D_e&\leq&\min_{a\in \hat{\Scal}}\Ebb[d(S, a)]
\end{eqnarray}
for some distribution $\Pbar_S\Pbar_{U|S}\Pbar_{X|SU}\Pbar_{YZ|X}$ and function $\phi(\cdot,\cdot)$.
\end{thm}
This trivial outer bound can be verified by using the optimality of hybrid coding for point-to-point communication and the fact that the estimation by the eavesdropper cannot be worse than the a-priori estimation of the source. 

\section{Numerical Example}
We use the same example that was considered in \cite{separate-schieler}. The source is distributed i.i.d. according to $Bern(p)$ and the channels are binary symmetric channels with crossover probabilities $p_1=0$ and $p_2=0.3$. For simplicity, we require lossless decoding at the legitimate receiver. Hamming distance is considered for distortion at the eavesdropper. 

\begin{figure}[H]
  \centering
  \includegraphics[width=9 cm]{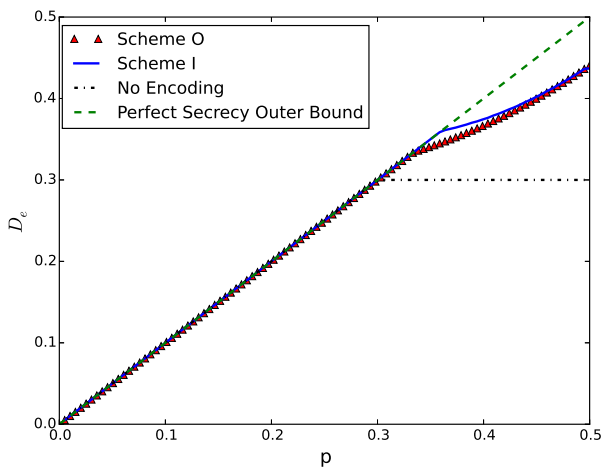}
\caption{Distortion at the eavesdropper as a function of source distribution $p$ with $p_1=0$, $p_2=0.3$.}
\label{compare}
\end{figure}

A numerical comparison of Scheme I with Scheme O is demonstrated in Fig. \ref{compare}. The choice of auxiliary random variable $U$ in Scheme I is $SX$, which may not necessarily be the optimum choice but is good enough to outperform Scheme O. Scheme II is not numerically evaluated. However, because of Corollary \ref{cor1} and \ref{cor2}, we know analytically that Scheme II is no worse than O or I.
\vspace{-0.4cm}
\vspace{-0.3in}
\section{Conclusion}
This work has investigated secure joint source-channel coding under a general information-theoretic secrecy formulation. 
By using hybrid coding, we achieve better performance than a previously considered operationally separate source-channel coding scheme (O). Although a simple numerical example shows that a basic hybrid coding scheme (I) can potentially outperform Scheme O, we have only managed to prove analytically a superposition hybrid coding scheme (II) can fully generalize both Scheme O and I. The direct relation between Scheme O and I, and whether Scheme II is strictly better than I are still open for further investigation. Non-trivial outer bounds are yet to be explored.
\section{Acknowledgement}
This research was supported in part by the Air Force Office of Scientific Research under Grant FA9550-12-1-0196 and MURI Grant FA9550-09-05086, in part by the Army Research Office under MURI Grant W911NF-11-1-0036, and in part by the National Science Foundation under Grants CCF-1116013, CCF-1350595, CNS-09-05086 and ECCS-1343210.
\bibliographystyle{ieeetr}
\bibliography{jscs}
\end{document}